\documentclass[%
 reprint, 
 amsmath,amssymb,
 aps, physrev,
]{revtex4-2}

\usepackage{dcolumn}
\usepackage{bm}
\usepackage{newtxtext,newtxmath}

\usepackage[T1]{fontenc}

\DeclareRobustCommand{\VAN}[3]{#2}
\let\VANthebibliography\thebibliography
\def\thebibliography{\DeclareRobustCommand{\VAN}[3]{##3}\VANthebibliography}
\def\refjnl#1{{\rmfamily#1}}%
\newcommand\apjl{\refjnl{ApJL}}     
\newcommand\apjs{\refjnl{ApJS}}
\newcommand\mnras{\refjnl{MNRAS}}
\newcommand\aap{\refjnl{A\&A}}
\newcommand\ssr{\refjnl{SSRv}}
\newcommand\nphysa{\refjnl{NuPhA}}

\newcommand{\nsat}{n_\mathrm{sat}}

\usepackage{graphicx}	
\usepackage{amsmath}	
\usepackage{xcolor}
\usepackage{isotope}

\usepackage{hyperref}

\begin{document}

\title[Testing Chiral EFT with RSFs]{Resonant shattering flares as asteroseismic tests of chiral effective field theory}

\author{Duncan Neill}%
 \email{Contact author: dn431@bath.ac.uk}
\author{David Tsang}
\email{D.Tsang@bath.ac.uk}
\affiliation{%
 Department of Physics, University of Bath, Claverton Down, Bath BA2 7AY, UK
}%

\author{Christian Drischler}
\email{drischler@ohio.edu}
\affiliation{%
 Department of Physics and Astronomy, Ohio University, Athens, Ohio 45701, USA
}%
\affiliation{%
Facility for Rare Isotope Beams, Michigan State University, East Lansing, Michigan 48824, USA
}%

\author{Jeremy W. Holt}
\email{holt@physics.tamu.edu}
\affiliation{%
 Cyclotron Institute, Texas A\&M University, College Station,
Texas 77843, USA
}%
\affiliation{%
Department of Physics and Astronomy,
Texas A\&M University, College Station, Texas 77843, USA
}%

\author{William G. Newton}
\email{William.Newton@tamuc.edu}
\affiliation{%
 Department of Physics and Astronomy, Texas A\&M University-Commerce, Commerce, Texas 75429-3011, USA
}%

\date{\today}

\begin{abstract}
Chiral effective field theory ($\chi$EFT) has proved to be a powerful microscopic framework for predicting the properties of neutron-rich nuclear matter with quantified theoretical uncertainties up to about twice the nuclear saturation density. Tests of $\chi$EFT predictions are typically performed at low densities using nuclear experiments, with neutron star (NS) constraints only being considered at high densities. In this work, we discuss how asteroseismic quasi-normal modes within NSs could be used to constrain specific matter properties at particular densities, not just the integrated quantities to which bulk NS observables are sensitive. We focus on the crust-core interface mode, showing that measuring this mode’s frequency would provide a meaningful test of $\chi$EFT at densities around half the saturation density. Conversely, we use nuclear matter properties predicted by $\chi$EFT to estimate that this mode’s frequency is around $185 \pm 50\,\text{Hz}$. Asteroseismic observables such as resonant phase shifts in gravitational-wave signals and multimessenger resonant shattering flare timings, therefore, have the potential to provide useful tests of $\chi$EFT.
\end{abstract}

\maketitle



\section{Introduction}

The nuclear matter equation of state (EOS) describes how the binding energy per nucleon of bulk nuclear matter varies with baryon number density, temperature (or equivalently energy/momentum), and isospin asymmetry (or likewise the proton fraction), making it a key component in modelling both nuclei \citep{Danielewicz2003Surface,Danielewicz2009Symmetry,Bombaci2018Equation} and neutron stars (NSs) \citep{steiner2008neutron,Lattimer2021Neutron}. 
However, the non-perturbative nature of the strong interaction at low energies makes an \textit{ab~initio} derivation of the EOS at densities relevant to nuclei and NSs very challenging, if not infeasible, at present. Various approaches have, therefore, been developed to calculate properties of nuclear matter based on nuclear forces fitted to two- and few-body observables.

Chiral effective field theory ($\chi$EFT), with Weinberg power counting, has become the dominant approach to deriving microscopic nuclear forces consistent with the symmetries (and symmetry-breaking pattern) of low-energy quantum chromodynamics (QCD).
It involves writing the most general Lagrangian consistent with the symmetries of low-energy QCD and using pion and nucleon effective degrees of freedom rather than QCD's quarks and gluons. Combined with a computational framework to solve the many-body Schr{\"o}dinger equation, $\chi$EFT can be used to calculate properties of nuclear matter with quantified uncertainties up to about twice nuclear saturation density, $\nsat \approx 0.16$\,fm$^{-3}$~\citep{Drischler:2024ebw}. By ordering the contributions to the nuclear potential in increasing powers of the ratio of a typical momentum scale of the system and the EFT breakdown scale, $\Lambda_b \approx 600$\,MeV, using a power counting scheme, one can obtain a high-fidelity description of the nucleon-nucleon and higher-body nuclear forces whose uncertainties due to omitted higher-order contributions can be estimated by analyzing the convergence in the chiral expansion. In recent years, there has been tremendous progress in applying $\chi$EFT to study the properties of atomic nuclei and bulk nuclear matter with rigorously quantified uncertainties~\citep{Hammer:2019poc,Tews:2020hgp,Hergert:2020bxy,Hebeler:2020ocj,Drischler:2021kxf,Hu:2021trw,Ekstrom:2022yea,Machleidt:2024bwl}. 

As with any theory, testing whether $\chi$EFT accurately calculates the properties of nuclear matter is important. Tests at low densities ($n \lesssim 2\nsat$), where $\chi$EFT is predictive, can offer insights into whether all the relevant low-energy physics has been included and $\chi$EFT works as advertised. Meanwhile, at high densities ($n \gtrsim 2\nsat$), where $\chi$EFT predicts its own breakdown (momentum) scale, tests can help in identifying the corresponding critical density and mechanism through which $\chi$EFT (in a particular implementation) breaks down in medium~\citep{2020DrischlerHowWell,Semposki:2024vnp}. Testing the predictions of $\chi$EFT for nuclear matter is therefore important over a wide range of densities, but currently there are only limited ways to do this validation~\citep{Koehn:2024set}.

At $n\lesssim 2\nsat$, the isospin-asymmetry dependence of the nuclear matter EOS expanded about isospin-symmetric matter is governed by the symmetry energy, the difference in binding energy between pure neutron matter and symmetric nuclear matter. Both the energy per particle of asymmetric matter and the symmetry energy can be calculated from $\chi$EFT \citep[see, e.g.,][and references therein]{Hebeler2015Nuclear,Sammarruca2019Nuclear,Drischler2021Towards} through the use of a many-body perturbation theory, and so constraints on them can be used to test the theory. While atomic nuclei are much closer to symmetric nuclear matter than pure neutron matter, NSs primarily consist of extremely neutron-rich matter, meaning that the symmetry energy holds particular importance for NSs. Being concerned with NSs, this work will therefore focus on the symmetry energy.

\begin{figure}
    \centering
    \includegraphics[width=\columnwidth,angle=0]{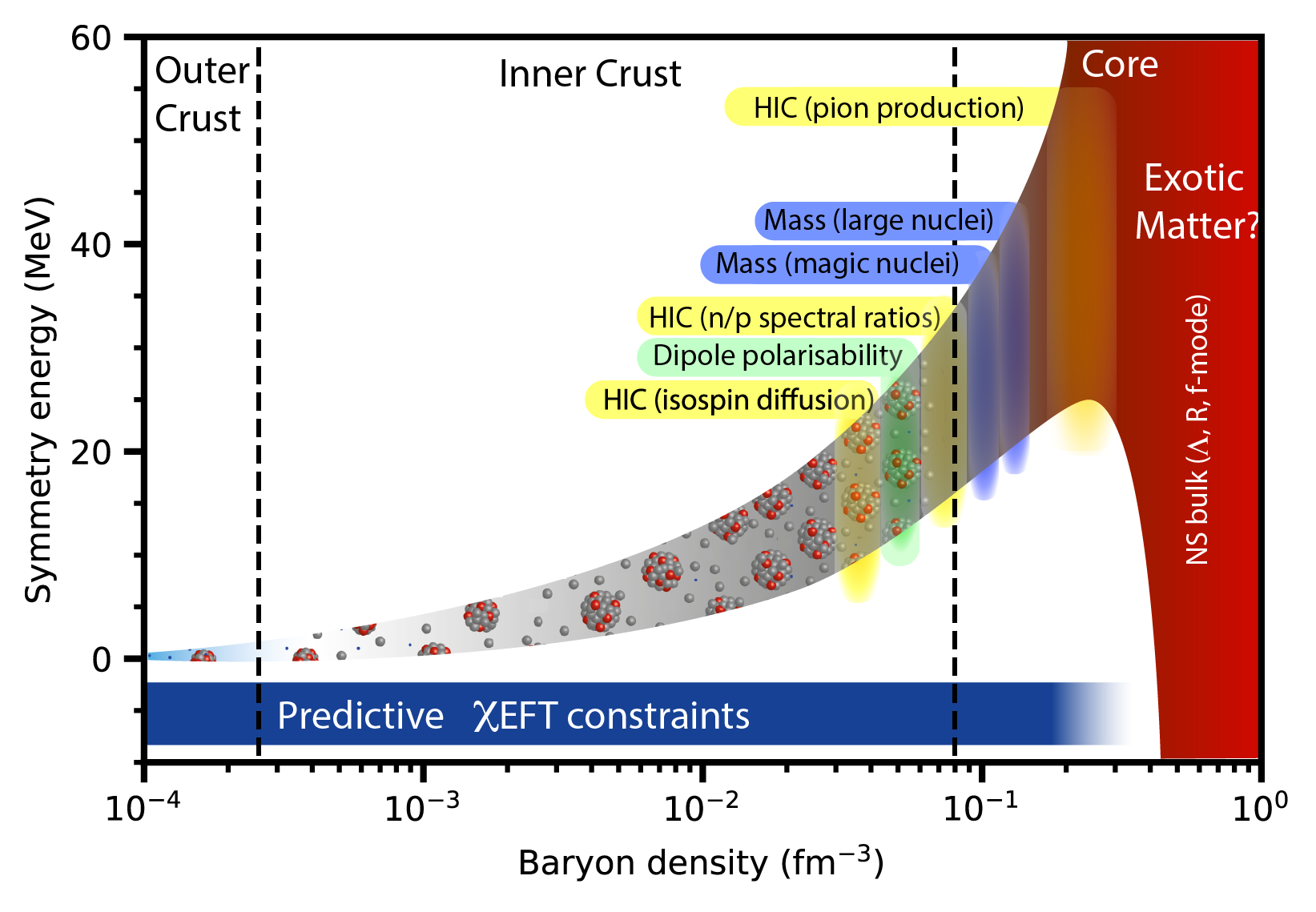}
    \caption{A schematic illustration of the densities at which various terrestrial experiments constrain the nuclear symmetry energy \citep{Lynch2022Decoding}, alongside broad bounds on what its value might be. The approximate densities at which $\chi$EFT is expected to be predictive for properties of nuclear matter are indicated by the blue line at the bottom. Labels along the top and dashed lines roughly split the density range by the regions of a NS, and the composition of NS matter in those regions is illustrated within the symmetry energy bounds. Bulk NS properties mainly inform us of the (possibly exotic) high-density matter found in the NS inner core, while experiment allows us to probe the symmetry energy at lower densities. Not shown here are experiments that can be used to probe the slope of the symmetry energy, such as the PREX and CREX experiments \citep{Adhikari2021PREXII,reed2021implications,Adhikari2022CREX,Reed_CREX2022}}
    \label{fig:cartoon}
\end{figure}

As illustrated in Figure~\ref{fig:cartoon}, much effort has been expended in a variety of terrestrial experiments to constrain the symmetry energy~\citep{Lynch2022Decoding}. For example, the products of heavy-ion collision (HIC) at various incident energies and impact parameters provide insight into the symmetry energy at different densities through distributions of their products and inferred nuclear matter transport dynamics \citep{tsang2009constraints,Morfouace2019Constraining,Estee2021Probing,Sorensen2024Dense}. Various projects to expand our knowledge of nuclear binding energies and charge radii also improve our understanding of the symmetry energy, as models fit to large or magic nuclei (i.e., nuclei with closed proton and neutron shells) display preferences for particular symmetry energy ranges \citep{Brown2013Constraints,Kortelainen2012Nuclear}. Recently, the \isotope[208]{Pb} and \isotope[48]{Ca} Radius EXperiments, PREX--II and CREX, have garnered particular interest as their results are difficult to reconcile with each other. These experiments measured parity-violating asymmetry in the elastic scattering of electrons from \isotope[208]{Pb} and \isotope[48]{Ca}~\citep{Adhikari2021PREXII,Adhikari2022CREX}, which can be used to obtain constraints on the thickness of the neutron skin of these nuclei -- a property with a strong correlation to the slope of the symmetry energy around $0.1\,\text{fm}^{-3}$ \citep{chen2010density,reed2021implications,Zhang2023CREX,Reed2024Density}. 

While these experiments provide symmetry energy constraints from various different phenomena, a commonality that many of them share is that they are primarily sensitive to the symmetry energy at densities below the saturation density. 
This is naturally a consequence of the density in nuclei peaking around saturation, meaning that nuclear matter must be compressed to probe its properties at higher densities. However, such compression requires high energies \citep[e.g.,][]{Estee2021Probing}, and thus momentum/temperature will have a significant effect on the nuclear matter EOS in these experiments. This means that terrestrial experiments leave high-density low-temperature parameter space for dense matter unexplored.

Instead, NSs are often put forward as ideal environments to probe the unexplored high-density and low-temperature region of the QCD phase diagram.
Their cores easily surpass the densities explored in laboratory experiments, reaching up to several times the saturation density, and old NSs will have cooled sufficiently to approximately consist of cold beta-equilibrium matter \citep{Nomoto1981Cooling,Lattimer1991Direct,Potekhin2015Neutron}. The use of NSs in studying high densities is supported by current observables -- such as x-ray pulse-profiles \citep{riley2019NICER,Miller2019NICER,riley2021NICER,Miller2021NICER}, relativistic Shapiro delay \citep{cromartie2020relativistic,Fonseca2021Refined}, and gravitational waves from coalescing binaries \citep{abbott2017gw170817} -- which provide constraints on bulk NS properties such as radius, mass (consequently constraining the maximum NS mass) and tidal deformability. These properties depend on the NS EOS, with particular sensitivity to the NS core and thus to the dense matter found there.
NSs with different masses will have different central densities, and so as the properties of more NSs are measured we will learn about matter over an increasingly broad high-density range.

When NSs are considered as sites to probe dense matter, it is therefore typically with focus on only the extreme densities found within their cores.
Of course, there is also low-density matter in the outer layers of NSs, and the NS inner crust and outer core maintain a high isospin-asymmetry that is ideal for studying the symmetry energy. However, bulk NS properties are largely insensitive to these outer layers or have a dependence on them that is degenerate with the core, making it difficult to reliably constrain low-density matter with sufficient precision to meaningfully aid experiment.
In this work, we will show that NS asteroseismic observables can provide insights into the NS crust with sufficient strength to aid terrestrial efforts to explore the subsaturation EOS and test the predictions of $\chi$EFT.

As different families of normal modes within NSs have properties determined by different facets of NS structure and composition, asteroseismology could be used to probe various properties of matter over a wider range of densities. Some, such as the gravitational ($g$-) and fundamental ($f$-) modes -- which may soon be probed through the resonant phase shifts they cause in gravitational waves (GWs) \citep{Andersson2018Using,Pratten2020Gravitational,Pratten2022Impact,Zhao:2022toc} -- are primarily sensitive to properties of the NS core, with the $f$-mode in particular giving information similar to that obtained from the tidal deformability \citep{Chan2014Multipolar,Exploring2020Andersson}. Other modes, however, are sensitive to details of the star's composition in less dense regions. For example, shear modes are restored by shear forces, and therefore have little dependence on matter outside the solid NS crust \citep{mcdermott1988nonradial}. The challenge of asteroseismology is not only in determining how different modes depend on the properties of nuclear matter, but also in identifying observables that contain signatures of modes, which is not an easy task given that the extreme mechanisms required for emissions to be observable at astronomical distances will typically be dominated by bulk NS properties.

A class of asteroseismic observables that may be detectable with current instrumentation are resonant shattering flares (RSFs). 
RSFs are brief gamma-ray flares triggered when the resonant excitation of a NS normal mode by the tidal field of its binary partner causes the NS crust to shatter \citep{tsang2012resonant,neill2022resonant}. As this is most likely to occur shortly before a binary merger when the tidal field is strongest, gravitational waves (GWs) could be detected coincident with RSFs.
Despite several steps separating the dynamics of the resonant mode from the emission of a flare, the multimessenger detection of GWs coincident with a RSF would allow for a simple and reliable measurement of that mode's frequency \citep{tsang2012resonant,neill2022resonant}.
The crust-core interface mode ($i$-mode) has been identified as a promising candidate for triggering RSFs, and its frequency is sensitive to the composition of the NS crust -- and thus to the EOS of nuclear matter -- near the crust-core transition \citep{neill2021resonant,Neill2023Constraining}, around half the nuclear saturation density. Measurements of the $i$-mode frequency from multimessenger RSF and GW events would, therefore, provide strong constraints on nuclear matter within NSs at that density, allowing for a low-density astrophysical test of $\chi$EFT.

RSFs and other asteroseismic observables may therefore expand the scope of NSs to being laboratories where nuclear matter can be probed at low densities as well as high densities, allowing for more comprehensive in-medium tests of $\chi$EFT. In this work, we will inject measurements of the $i$-mode frequency into inferences of the nuclear symmetry energy via a NS meta-model, which takes extended Skyrme models for the nuclear matter EOS as inputs to construct models for the NS EOS and composition. We will show that the resulting posteriors indicate that such measurements would be useful as tests of $\chi$EFT by comparing them to conservative $\chi$EFT predictions for the nuclear symmetry energy, and that when used in conjunction with experimental constraints on the symmetry energy, they can improve tests to keep pace with optimistic predictions for $\chi$EFT uncertainties.

\section{Constraining the symmetry energy}\label{sec:Esym_constraints}

\subsection{Neutron star meta-model and prior}\label{sec:model_prior}
We parameterise the equation of state (EOS) of bulk nuclear matter using an extended Skyrme interaction~\citep{Lim:2017aa} that allows for independent variation of the first three parameters of the expansions of the symmetric matter EOS (energy per particle $E_0$, saturation density $\nsat$, and incompressibility $K_0$) and symmetry energy (symmetry energy at $\nsat$ $J$, slope parameter $L$, and the second-order coefficient $K_{\rm sym}$) around the nuclear saturation density. We will however fix the symmetric matter parameters to $E_0=-15.93\text{ MeV}$, $\nsat=0.1562\text{ fm}^{-3}$, and $K_0=239.5\text{ MeV}$ \citep[values based on the Sk$\chi$450 parametrisation of][]{Lim:2017aa}, to focus on the symmetry energy. 
We use this Skyrme model for simplicity, as it has been extensively used in previous studies of the $i$-mode and its connection to the properties of nuclear matter. More detailed work in the future might instead sample the EOS from the predictions of $\chi$EFT to make the steps involved in modelling NSs more consistent, but that is unnecessary for this first investigation.

The extended Skyrme interaction is used with the compressible liquid drop model to calculate the EOS and composition of the solid NS crust, and is used in the fluid outer core to consistently calculate the properties of uniform matter. At $1.5\nsat$, however, we switch to a polytropic model with a piecewise transition at $2.7\nsat$, allowing the inner core to deviate from nucleonic matter to represent our uncertainty in the nature of matter there. The indices of these polytropes ($\gamma_1$ and $\gamma_2$) can be varied alongside the symmetry energy parameters, giving us a five-parameter NS meta-model (a model for constructing NS models). For further details of this meta-model, see \citet{Newton:2013aa,balliet2021prior,neill2021resonant}.

We construct a broad prior for these five meta-model parameters, beginning by selecting ranges for the symmetry energy parameters that are sufficient to confidently cover values inferred from a wide range of nuclear experiment and theory \citep[see for example][and references therein]{Margueron2018Equation}: $25<J<40\text{ MeV}$, $0<L<160\text{ MeV}$, and $-500<K_{\rm sym}<200\text{ MeV}$. Aside from values outside these ranges, any values within them that produce nonphysical EOSs when used in our Skyrme model (e.g., by having stable pure neutron matter below $1.5\nsat$) will also have zero probability in our prior. 
Meanwhile, to ensure that we include all parameter values for which NS matter remains stable and causal ($0 \leq c_s \leq 1$, where $c_s$ is sound speed in units of the speed of light), we use $1<\gamma_1<7$ and $1<\gamma_1<7$. These ranges are extremely broad, and so as part of our prior we also include a constraint on the NS maximum mass, which is mainly determined by these parameters. NSs have been identified with masses up to $\approx 2.1\text{ M}_{\odot}$ \citep{Antoniadis2013Massive,Fonseca2021Refined}, but to keep our prior broad and avoid overly constraining the NS core due to the limited flexibility of our two-piece poltropic core model, we use the conservative constraint $1.9\leq M_{\rm max}\leq 2.6\text{ M}_{\odot}$. 
Following the discussion of \citet{Neill2024Strengthening}, we then use a prior weighting that varies with the symmetry energy parameters to produce a probability distribution that is uniform in the 3D $J$-$L$-$K_{\rm sym}$ space obtained when marginalising over $\gamma_1$ and $\gamma_2$, which avoids the bias in the symmetry energy that a simpler uniform 5D prior would produce due to our choice to use polytropes in the core.

For real multimessenger RSF and GW events, there will also be uncertainty in the mass of the NS that produced the RSF, which will affect the $i$-mode frequency. However, as shown in \citet{Neill2024Strengthening}, the $i$-mode is only weakly dependent on NS mass, and therefore we do not concern ourselves with constructing a highly realistic mass prior for NS merging binaries. Instead we simply use a broad uniform probability distribution between $0.9\text{ M}_{\odot}$ and $1.9\text{ M}_{\odot}$.

\subsection{Injecting $i$-mode frequency measurements and comparison to $\chi$EFT}

For simplicity, we shall assume that measurements of the $i$-mode frequency give likelihood functions that are normal distributions with standard deviations of $15\text{ Hz}$. We will discuss the possibility of obtaining a measurement of this strength from multimessenger RSF and GW events in Section~\ref{sec:discuss}, but such precision is not unrealistic \citep{neill2021resonant,Neill2023Constraining}. 
The $i$-mode frequency for a chosen NS model and mass can be calculated by solving the Tolman-Oppenheimer-Volkoff equations and relativistic pulsation equations, the latter for which we follow \citet{yoshida2002nonradial} (which applies the relativistic Cowling approximation). This requires the NS EOS and information about the composition of the NS in the form of the shear modulus and the adiabatic index for matter with frozen composition. We choose to have the centres of our injected likelihood functions be values close to the ends of the $i$-mode frequency distribution of samples drawn from our prior, to emphasise the difference in constraints from different frequency values: $140\text{ Hz}$ and $410\text{ Hz}$. For the NS mass meanwhile, we will always inject a value of $1.4\text{ M}_{\odot}$, multiplying our likelihood functions by a normal distribution centered on that value and with a deviation of $0.1\text{ M}_{\odot}$, roughly based on the scale of mass uncertainties that can be obtained from GWs \citep{Neill2024Strengthening}.

\begin{figure}
    \centering
    \includegraphics[width=\columnwidth,angle=0]{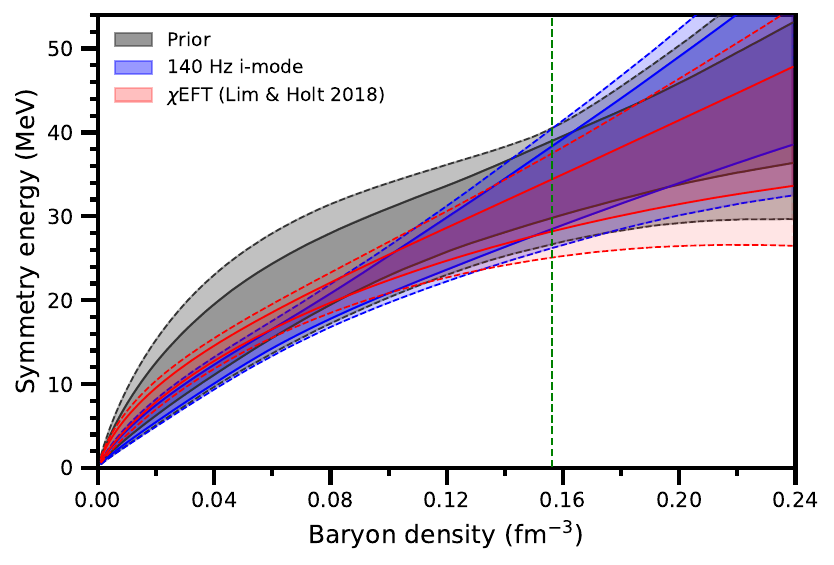}
    \includegraphics[width=\columnwidth,angle=0]{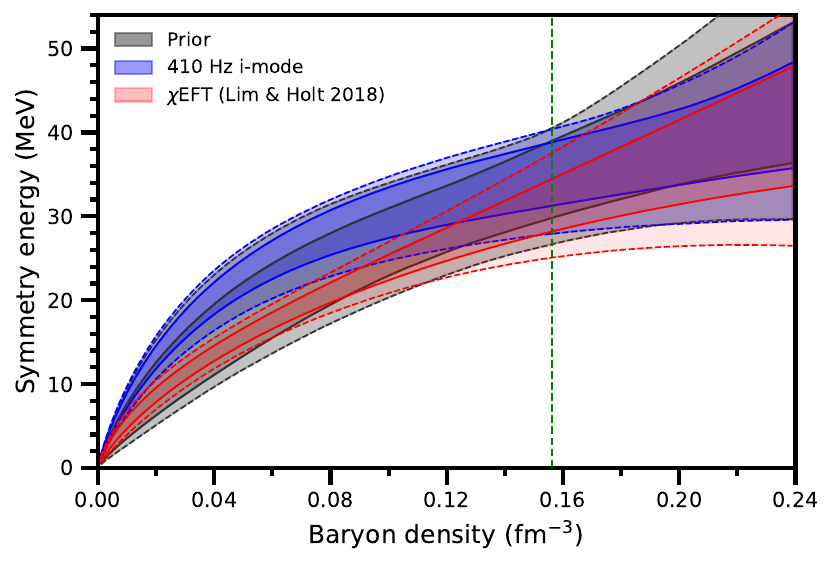}
    \caption{Prior (grey) and posterior (blue) constraints on the nuclear symmetry energy over the range of densities used by our meta-model, showing what could be learned from measurements of the $i$-mode frequency. The top panel is for an injected $i$-mode frequency measurement of $140\text{ Hz}$, and the bottom panel for $410\text{ Hz}$. Also shown (in red) are conservative constraints from $\chi$EFT \citet{lim2018neutron}. Solid (dashed) lines bound the central 68\% (95\%) credibility regions. The vertical dashed line indicates the nuclear saturation density used in our NS meta-model (and is unrelated to the $\chi$EFT constraints).}
    \label{fig:Esym_viable_LimHolt}
\end{figure}

In Figure~\ref{fig:Esym_viable_LimHolt}, we show contours for the symmetry energy as a function of density containing 68\% and 95\% of our inferred prior samples (in grey) and posterior samples for $140\text{ Hz}$ and $410\text{ Hz}$ injected $i$-mode frequency measurements (in blue).
These posteriors clearly show that while the $i$-mode provides little information about the symmetry energy above saturation density, it is informative below half saturation, as the posteriors in the two plots prefer different values there. 

To compare these posteriors to predictions of $\chi$EFT, we consider the symmetry energy priors of \citet{lim2018neutron}. These are obtained by sampling two multivariate probability distributions for parameters in a nuclear matter energy density functional -- one for parameters describing the energy density of pure nuclear matter, and the other for symmetric nuclear matter -- to reproduce a range of $\chi$EFT calculations using various resolution scales, chiral orders, and many-body perturbation theory orders. This method is relatively conservative in its quantification of uncertainties in $\chi$EFT, and therefore represents a conservative bound on the symmetry energy constraints that can currently be obtained from $\chi$EFT.

We plot these $\chi$EFT constraints in red alongside our prior and posteriors in Figure~\ref{fig:Esym_viable_LimHolt}. Our prior contains the range of possibilities from $\chi$EFT reasonably well, and in particular is much broader at the sub-saturation densities to which the $i$-mode is sensitive. This indicates that it does not contain significant information not also considered by $\chi$EFT, making it a reasonable starting point from which to examine how the $i$-mode could be used to test $\chi$EFT. By examining the posteriors for the different $i$-mode frequencies we see that such tests are indeed promising, as at low densities the posteriors are constrained significantly compared to the prior and have uncertainties that are not large when compared to those in $\chi$EFT's predictions, meaning that there can be clear agreement or conflict between the $i$-mode frequency measurements and $\chi$EFT, depending on the measured frequency.

\begin{figure}
    \centering
    \includegraphics[width=\columnwidth,angle=0]{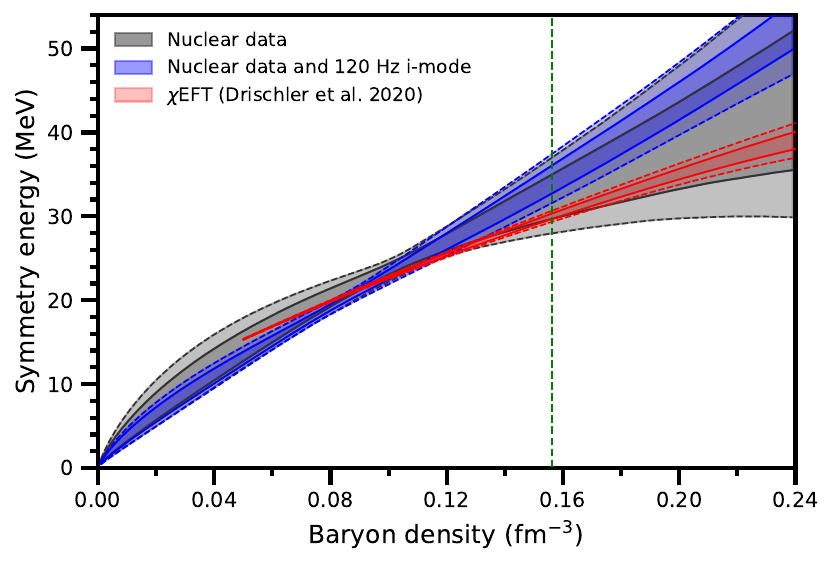}
    \includegraphics[width=\columnwidth,angle=0]{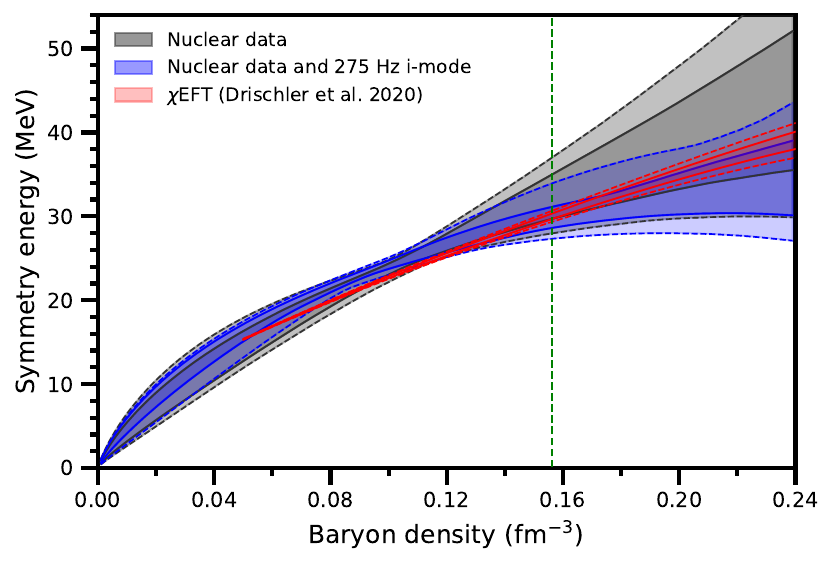}
    \caption{Similar to Figure~\ref{fig:Esym_viable_LimHolt}, but for posteriors informed by various experimental nuclear data (see Table~2 of \citet{Neill2024Strengthening}). The grey constraints are for this data alone, while the blue ones also include injected $i$-mode frequency measurements: at $120\text{ Hz}$ for the top panel, and at $275\text{ Hz}$ for the bottom panel. Even with the addition of experimental data, $i$-mode frequency measurements clearly still improve symmetry energy constraints below saturation and thus could help to test $\chi$EFT there.}
    \label{fig:Eym_nucler_Drischler}
\end{figure}

Measurements of the $i$-mode frequency could, therefore, be useful in testing conservative $\chi$EFT predictions for the symmetry energy below the saturation density. However, these measurements would be even more useful if they could be used to test less conservative predictions for $\chi$EFT uncertainties in the symmetry energy at low density. Considering the state-of-the-art $\chi$EFT symmetry energy predictions of \citet{2020DrischlerQuantifying} \citep[see also][]{2019DrischlerChiral,2020DrischlerHowWell}  (shown in red in Figure~\ref{fig:Eym_nucler_Drischler}) -- which make use of correlations between the uncertainties in the binding energies of pure neutron and symmetric nuclear matter predicted by next-to-next-to-next-to-leading order (N$^3$LO) $\chi$EFT to obtain stronger symmetry energy constraints -- we see that they are much more tightly constrained than our posteriors. Therefore, a measurement of the $i$-mode frequency with the precision assumed here will only provide a very general test of these $\chi$EFT constraints.

However, constraints on the symmetry energy provided by the $i$-mode frequency and terrestrial observables are complementary \citep{Neill2023Constraining}, meaning that by considering the consistency of $\chi$EFT with measurements of the $i$-mode frequency and other observables simultaneously, we may be able to test more less conservative predictions. Examining current constraints from experimental nuclear physics, some of the more well-determined quantities are the binding energies \citep{Wang2021AME} and charge radii \citep{Angeli2013Table} of doubly magic nuclei, which are sensitive to the symmetry energy at similar densities as the $i$-mode \citep[][]{kortelainen2010nuclear,Brown2013Constraints,Kortelainen2012Nuclear}. Therefore, we use the Skyrme model underlying our NS meta-model to calculate these properties for sampled $J$, $L$, and $K_{\rm sym}$ values and construct a likelihood that compares them to experimental values. The data we use are the same binding energies and charge radii as listed in Table~2 of \citet{Neill2024Strengthening} (excluding the neutron skins or dipole polarisabilities listed there), and following that work, we construct Gaussian likelihood functions for each observable, taking their product as an overall experimental nuclear likelihood. Using only this likelihood to infer the symmetry energy parameters, we obtain the set of posteriors shown in grey in Figure~\ref{fig:Eym_nucler_Drischler}. These posteriors are consistent with $\chi$EFT, and the symmetry energy is most constrained around $(2/3)\nsat$.

Combining the experimental nuclear likelihood with likelihoods for injected $i$-mode frequency measurements at $120\text{ Hz}$ and $275\text{ Hz}$ (values chosen for being near the ends of the range that is consistent with the experimental symmetry energy posteriors), we obtain the posteriors show in blue in Figure~\ref{fig:Eym_nucler_Drischler}. The symmetry energy constraints around half the saturation density are significantly improved by the injections, and the values each injection prefers are noticeably different, which again indicates a correlation between the $i$-mode frequency and the symmetry energy at these densities.
Compared to the $\chi$EFT uncertainties of \citet{2020DrischlerQuantifying}, these combined posteriors are not very broad, making this combination of data more meaningful as a test of $\chi$EFT than the $i$-mode frequency is alone. They also indicate that, even when accounting for nuclear observables that constrain the symmetry energy at similar densities, the $i$-mode frequency contains new information useful for assessing $\chi$EFT predictions.

\section{$\chi$EFT predictions for the $i$-mode frequency}\label{sec:predict_freq}

So far, we have taken $\chi$EFT predictions for the symmetry energy and confronted them with values for the symmetry energy inferred from the $i$-mode frequency. By instead examining how the symmetry energy is correlated with the $i$-mode frequency, we can use $\chi$EFT predictions for the symmetry energy to predict the $i$-mode frequency.

\begin{figure}
    \centering
    \includegraphics[width=\columnwidth,angle=0]{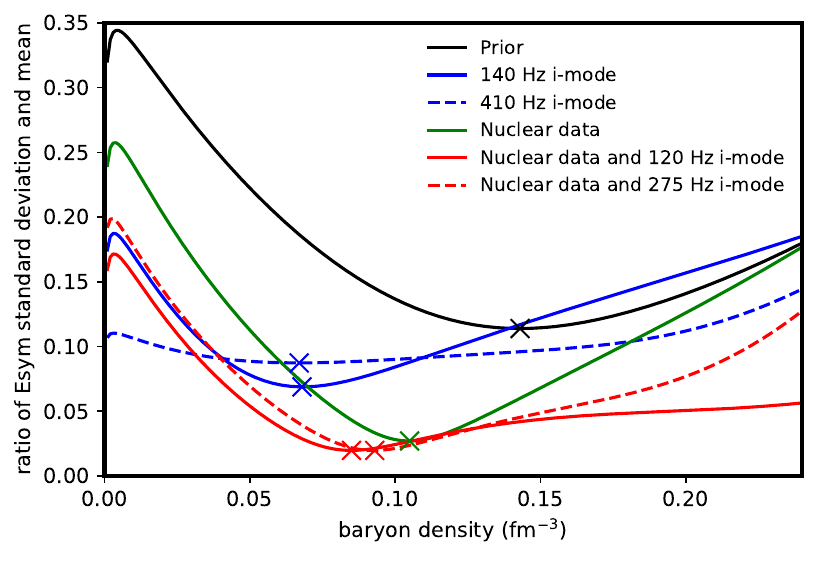}
    \caption{The ratio of the standard deviation and the mean of the symmetry energy of samples drawn from our priors and posteriors for injected $i$-mode measurements and nuclear masses. The marker on a line at its minimum indicates the density at which the symmetry energy is the most constrained in that inference, and thus, the density where the quantity used as data is most sensitive to the symmetry energy.}
    \label{fig:constraint_width}
\end{figure}

We begin by investigating the density at which the value of the symmetry energy has the strongest correlation with the $i$-mode frequency. To do this, we examine how the standard deviations of the symmetry energies of our samples vary with density. This examination is shown in Figure~\ref{fig:constraint_width}, where we have taken the ratio of the standard deviation and the mean symmetry energy to account for the way in which our model has the samples converge to zero symmetry energy at zero density. We see that the posteriors informed only by the $i$-mode frequency have the symmetry energy most constrained around $n \approx 0.068\text{ fm}^{-3}$, while those informed by nuclear masses primarily constrain the symmetry energy around $n \approx 0.105\text{ fm}^{-3}$. This finding indicates a sensitivity of these quantities to those densities.

\begin{figure*}
    \centering
    \includegraphics[width=\textwidth,angle=0]{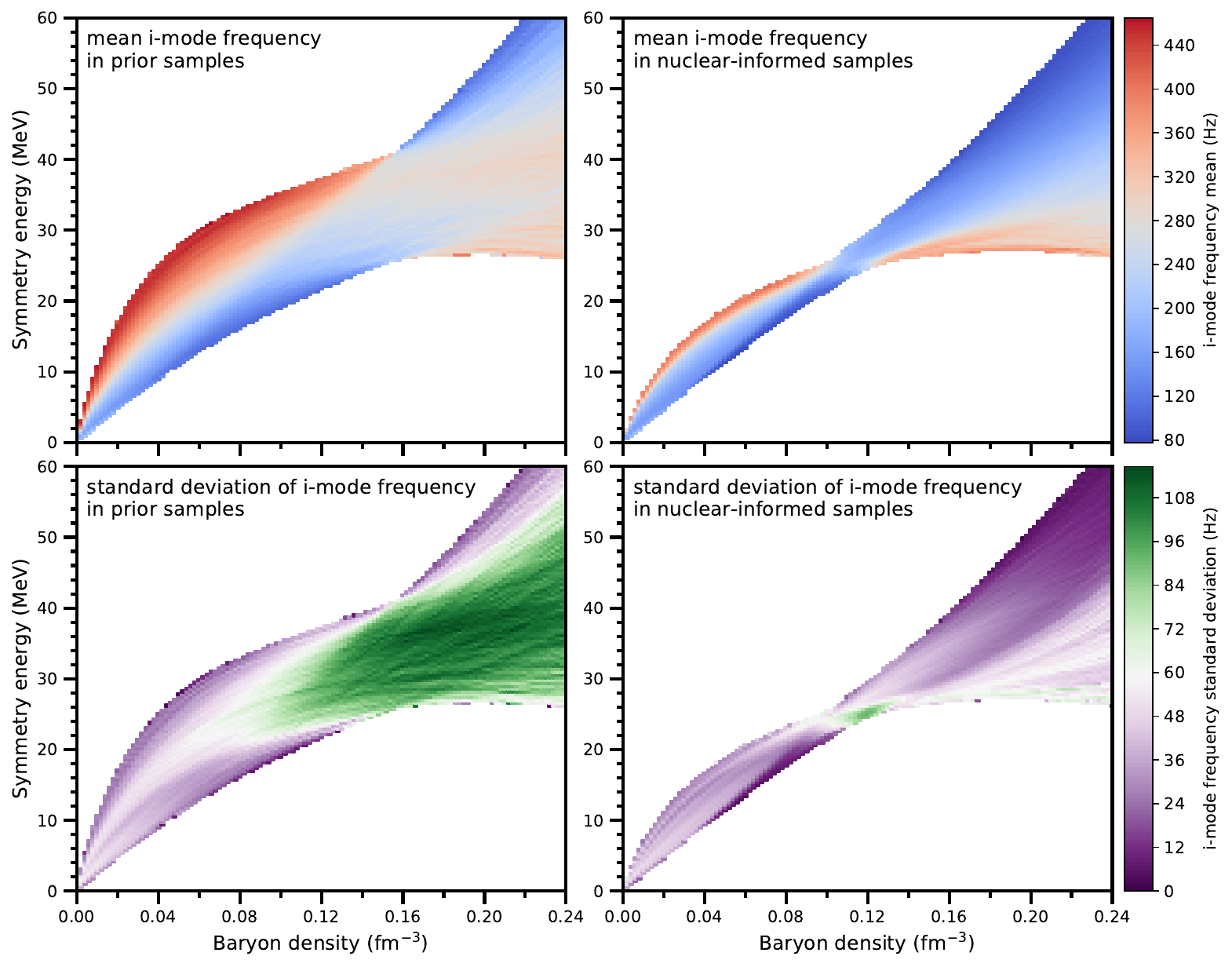}
    \caption{The mean (top panels) and standard deviation (bottom panels) $i$-mode frequency of sampled EOSs that fall into bins in symmetry energy at various densities. The plots on the left use samples are taken from our prior, and the ones on the right use samples from our nuclear-informed posteriors. There is a correlation between the mean $i$-mode frequency and the symmetry energy below $\sim0.1\text{ fm}^{-3}$, and the relatively low standard deviation there indicates that it is a strong correlation.}
    \label{fig:Esym_bins_f2i}
\end{figure*}

We now examine the relationship between the symmetry energy and $i$-mode frequency. To display this relationship and how it differs when considering the symmetry energy at different densities, we bin the samples in our prior by their symmetry energies at a given density, and then calculate the mean and standard deviation of the $i$-mode frequencies of the samples in each bin. Repeating this over the nuclear matter EOS density range used in our NS meta-model, and repeating the whole process for our nuclear-informed posteriors, we obtain Figure~\ref{fig:Esym_bins_f2i}. From this Figure we see that the symmetry energy below approximately half saturation density has a clear linear correlation with the $i$-mode frequency, with a relatively low standard deviation that indicates that the correlation is strong. Following Figure~\ref{fig:constraint_width} and focusing more specifically on the correlation around $0.068\text{ fm}^{-3}$, the range of $i$-mode frequencies spanned by the $\chi$EFT symmetry energy predictions of \citet{2020DrischlerQuantifying} shown in Figure~\ref{fig:Eym_nucler_Drischler} suggests that the $i$-mode frequency is around $185\pm 50\text{ Hz}$, where the uncertainty almost entirely comes from the standard deviation of the $i$-mode frequency in the bins, not from the uncertainty in the $\chi$EFT symmetry energy predictions. This is generally supported by Figures~\ref{fig:Esym_viable_LimHolt} and~\ref{fig:Eym_nucler_Drischler}, where the posteriors for frequency measurements injected at $140$ and $410\text{ Hz}$, and at $120$ and $275\text{ Hz}$, bracket the $\chi$EFT predictions.

\section{Discussion}\label{sec:discuss}

The values of the $i$-mode frequency given in this work -- including the prediction of $185\pm50\text{ Hz}$ based on $\chi$EFT -- might change significantly if improvements are made to the NS model, particularly if they affect the composition of the NS crust. For example, we have used the compressible liquid drop model to describe the clustering of nucleons in the crust, but more complex models may result in different behaviours. Of particular importance would be the inclusion of nuclear pastas: exotic nuclear geometries that appear near the base of the NS crust. Different phases of pasta could have very different shear moduli \citep{Pethick:1998aa,Caplan2018Elasticity,Pethick2020Elastic}, altering the main force that restores the $i$-mode and therefore changing the frequency calculated for any given nuclear matter EOS. A measurement of the $i$-mode frequency that is inconsistent with our prediction could therefore indicate that our NS modeling is not accurate, rather than that there is a problem in $\chi$EFT. Improving our understanding of the structure of nuclear matter in the NS crust is therefore required in order for this asteroseismic test of $\chi$EFT to be reliable.

In this work, we have examined whether the $i$-mode frequency would be useful as a test of $\chi$EFT if it were measured with an uncertainty on the order of $15\text{ Hz}$. While we have found that this is the case, since we currently have no $i$-mode frequency measurements, there is a question of whether this uncertainty is reasonable. The uncertainty in a measurement of the $i$-mode frequency from multimessenger GW and RSF events primarily comes from the range of frequencies the GWs sweep through over the duration of the RSF~\citep{neill2021resonant}. As the rate at which GW frequency changes increases with frequency and during resonance the GW frequency equals the $i$-mode frequency, the uncertainty in the $i$-mode frequency will be larger the higher its real value is. Following~\citet{neill2021resonant}, the uncertainty in $i$-mode frequency measurements will be $15\text{ Hz}$ if the $i$-mode frequency is around $200\text{ Hz}$, meaning that for frequencies consistent with $\chi$EFT the uncertainty we have used here is reasonable.

If the $i$-mode actually has a higher frequency, an uncertainty of $15\text{ Hz}$ could still be achieved by combining measurements of the $i$-mode frequency from different RSFs \citep{Neill2024Strengthening}. Whether this is possible will, however, depend on the rate at which multimessenger RSF and GW detections occur. On the GW side, current interferometers are pushing out the detection threshold for binary NS mergers to distances similar to those at which RSFs might be detectable with current telescopes \citep{neill2022resonant}. With aLIGO reaching design sensitivity in the near future and new interferometers being planned, we may soon be in a position to detect the majority of nearby mergers. The rate at which RSFs occur and are close enough to be observed is however uncertain, as no observed EM transients have yet been confirmed as RSFs. However, short gamma-ray burst precursor flares might be RSFs \citep{tsang2012resonant}, and around $\sim1$ of these are detected each year \citep{troja2010precursors,Minaev2017Precursors,zhong2019precursors,Wang2020Stringent,Li2021Temporal}. If a significant fraction of these precursors are RSFs, we might expect to observe multiple multimessenger events in the near future.

RSFs are not the only asteroseismic observable that could provide insight into the composition of the NS crust. For example, quasi-periodic oscillations (QPOs) in giant flares from soft gamma repeaters may be signatures of torsional shear mode oscillations~\citep{Duncan1998Global,Israel2005Discovery,Strohmayer2005Discovery,Watts2006Detection}, in which case matching the QPO frequencies to those modes in NS models could provide a probe of the crust. Also, as next-generation GW interferometers will significantly improve the signal-to-noise in detected GW transients, the need for an EM flare to indicate the timing of $i$-mode resonance could fade away, as the GW phase shifts caused by mode resonances may be detectable and thus allow for $f$-, $g$- and $i$-modes to be detected from GWs alone \citep{Andersson2018Using,Pratten2020Gravitational,Pratten2022Impact}. Even if detectable RSFs are extremely rare, we may, therefore, eventually be able to sufficiently constrain the $i$-mode frequency from GWs alone to provide a test of $\chi$EFT. Finally, we note that low-order $g$-modes with similar eigenfunctions to the $i$-mode \citep{Passamonti2021Dynamical,Gittins2024Neutron} have also been suggested as candidates for triggering RSFs \citep{Kuan2021General1,Kuan2021General2}, but that will depend on whether these modes are strongly suppressed within a solid and stratified NS crust \citep{mcdermott1988nonradial}. Overall, as new techniques are developed and instruments constructed, asteroseismic probes may become more commonplace and give more detailed insight into NS composition, greatly enhancing our ability to probe dense matter in these compact objects.

\subsection{Caveats and model dependences}

The EOS of symmetric nuclear matter being fixed in our inferences creates some ambiguity in how to compare the resulting posteriors to $\chi$EFT predictions, as it means that our posteriors for the EOS of pure neutron matter and the symmetry energy contain the same information, which is not the case for those from $\chi$EFT \citep[particularly for those in][]{2020DrischlerQuantifying}. However, for a model with more freedom in the symmetric matter EOS, we would expect measurements of the $i$-mode frequency to mainly inform us of the symmetry energy and not the EOS of pure neutron matter. This is because the $i$-mode is sensitive to the EOS of nuclear matter via the proton fraction of the crust \citep[through how it affects the shear modulus][]{neill2021resonant}, and the proton fraction is determined by the symmetry energy, not the EOS of pure neutron matter. It is more appropriate to compare to $\chi$EFT's predictions for symmetry energy rather than for pure neutron matter. If the parameters of symmetric matter were allowed to vary, we would expect the symmetry energy posteriors shown in Figures~\ref{fig:Esym_viable_LimHolt} and~\ref{fig:Eym_nucler_Drischler} to be largely unchanged, while the posteriors on pure neutron matter would shift by an amount strongly correlated with the uncertainty in symmetric matter.

All of the $i$-mode-frequency-informed posteriors in Figures~\ref{fig:Esym_viable_LimHolt} and ~\ref{fig:Eym_nucler_Drischler} show improvement in symmetry energy constraints above the saturation density, as well as below it. However, this is a model-dependent feature, resulting from the extended Skyrme model used in our NS meta-model having only three free parameters with which to vary the symmetry energy, which limits the flexibility of how the symmetry energy can vary with density. We identify the change in the posteriors below saturation as a direct consequence of the $i$-mode frequency data and the change above it as model-dependent because the $i$-mode is restored by shear forces -- which, as mentioned above, are dependent on the symmetry energy -- and exists due to the discontinuous transition between the NS crust and core. As the crust-core transition occurs around $n \approx 0.08-0.10\text{ fm}^{-3}$ and the fluid core does not support shear, there is therefore a clear reason for the $i$-mode to be sensitive to the symmetry energy at and below that density. There is, however, no such physical explanation for a dependence on the symmetry energy above saturation density, which is far into the fluid core.

Another spurious correlation in this work is between the curvature of the symmetry energy below the saturation density and the $i$-mode frequency, which can be seen in Figure~\ref{fig:Esym_bins_f2i} from how the curvature of the red and blue regions differ. However, as justified above, the dynamics of the $i$-mode are connected to the symmetry energy and not its curvature, so the correlation with the latter is not physical. Instead, this correlation appears in our prior due to our choice to restrict the symmetry energy at the saturation density to the range $25<J<40\text{ MeV}$. EOSs with high symmetry energies below the saturation density must, therefore, have more negative curvature to pass between these bounds while keeping the symmetry energy fixed to zero at zero density, and vice-versa for EOSs with low symmetry energy. The dependence of the $i$-mode frequency on symmetry energy then combines with the correlation between symmetry energy and its curvature to correlate the frequency and curvature. A similar correlation also appears in our nuclear-informed posteriors due to the nuclear data ``pinching'' the symmetry energy around $n\approx 0.10\text{ fm}^{-3}$. As with the other model-dependent correlations discussed here, we expect that this would disappear if we were to use a model which allowed more freedom in how the symmetry energy varies with density.

The details of this work are therefore model dependent, and caution must be taken when analysing our results. When real $i$-mode frequency data is obtained and quantitatively meaningful results are needed, the modeling described in this work could be improved, such as by using EOSs sampled from $\chi$EFT predictions in a NS meta-model rather than the Skyrme model, which could help reduce spurious correlations along the EOS. However, for the purpose of this work, the methods we have used are sufficient to obtain a reasonable qualitative conclusion: asteroseismic quantities such as the $i$-mode frequency can provide insight into the properties of low-density neutron star matter that is sufficient to be used alongside experimental constraints in tests of $\chi$EFT.

\section{Conclusions}

In this work, we have examined how measurements of the neutron star crust-core interface mode's frequency, which could be obtained from coincident resonant shattering flare and gravitational wave detection, can be used to probe the nuclear matter equation of state at sub-saturation densities. We have shown that the $i$-mode frequency is sensitive to the symmetry energy below half the saturation density, and that measurements of it with realistic uncertainties constrain the symmetry energy sufficiently to provide useful tests of chiral effective field theory. Within the limitations of our neutron star meta-model, we found that $\chi$EFT predicts $i$-mode frequencies of around $185\pm50\text{ Hz}$. 

Neutron stars are, therefore, not only laboratories in which the nuclear matter EOS can be studied at high densities but also at densities below saturation. Combined with constraints already available from nuclear experiment, the $i$-mode provides new insights that can help assess the validity of predictions made using $\chi$EFT. More broadly, constraints on the $i$-mode through resonant shattering flares are an example of how asteroseismic observables can offer powerful probes of neutron star matter that allow for targeted study of specific properties and densities of nuclear matter.

\section*{Acknowledgements}
DN and DT were supported by the UK Science and Technology Facilities Council (ST/X001067/1) and the Royal Society (RGS/R1/231499). 
The work of J.W.H.\ is supported by the U.S.\ National Science Foundation under grant PHY-2209318.
C. D. acknowledges support from the National Science Foundation under award PHY~2339043.
WGN was supported by NASA award 80NSSC18K1019 and the National Science Foundation under award PHY-2209536.
This material is based upon work supported by the U.S. Department of Energy, Office of Science, Office of Nuclear Physics, under the FRIB Theory Alliance award DE-SC0013617.


\bibliographystyle{mnras}
\bibliography{RSFSymmetry}

\label{lastpage}
\end{document}